\documentclass[aps,prl,superscriptaddress,preprintnumbers,twocolumn,amsmath,amssymb,floatfix]{revtex4-1}
\usepackage{graphicx}
\usepackage{hyperref}
\usepackage{color}
\definecolor{RED}{rgb}{1,0,0}
\definecolor{BLUE}{rgb}{0,0,1}
\begin{document}

\title{Tracing  crystal-field splittings in the rare earth-based intermetallic  CeIrIn$_5$}
\author{Q. Y. Chen}
\affiliation{State Key Laboratory of Surface Physics and Department of Physics, Fudan University, Shanghai 200433, China}
\affiliation{Science and Technology on Surface Physics and Chemistry Laboratory, Mianyang 621908, China}
\author{C. H. P. Wen}
\author{Q. Yao}
\author{K. Huang}
\author{Z. F. Ding}
\author{L. Shu}
\author{X. H. Niu}
\affiliation{State Key Laboratory of Surface Physics and Department of Physics, Fudan University, Shanghai 200433, China}
\author{Y. Zhang}
\author{X. C. Lai}
\affiliation{Science and Technology on Surface Physics and Chemistry Laboratory, Mianyang 621908, China}
\author{Y. B. Huang}
\affiliation{Shanghai Institute of Applied Physics, CAS, Shanghai, 201204, China}
\author{G. B. Zhang}
\affiliation{National Synchrotron Radiation Laboratory, University of Science and Technology of China, Hefei, 230029, China}
\author{S. Kirchner}
\email{stefan.kirchner@correlated-matter.com}
\affiliation{Center for Correlated Matter, Zhejiang University, Hangzhou, 310058, China}
\author{D. L. Feng}
\email{dlfeng@fudan.edu.cn}
\affiliation{State Key Laboratory of Surface Physics and Department of Physics, Fudan University, Shanghai 200433, China}
\affiliation{Collaborative Innovation Center of Advanced Microstructures, Nanjing 210093, China}

\begin{abstract}
Crystal electric field states in rare earth intermetallics show an intricate entanglement with the many-body physics that occurs in these systems and that is known  to lead to a plethora of electronic phases. Here, we attempt to trace  different contributions to the crystal electric field (CEF) splittings in CeIrIn$_5$, a heavy-fermion compound  and member of the Ce$M$In$_5$ ($M$= Co, Rh, Ir) family. To this end, we  utilize  high-resolution resonant angle-resolved photoemission spectroscopy (ARPES) and present a spectroscopic study of the electronic structure of this unconventional superconductor over a wide  temperature range.   As a result, we show how ARPES can be used in combination with  thermodynamic measurements or neutron scattering to disentangle different contributions to the CEF splitting in rare earth intermetallics.
We also find that the hybridization is stronger in CeIrIn$_5$ than CeCoIn$_5$ and the effects of the hybridization on the Fermi volume increase is much smaller than predicted.
By providing the first experimental evidence  for  $4f_{7/2}^{1}$ splittings which, in CeIrIn$_5$,  split the octet into four doublets, we clearly demonstrate the many-body origin of the so-called $4f_{7/2}^{1}$ state.

\end{abstract}

\maketitle
\section*{I. Introduction}
Systems possessing a large spin-orbit interaction have recently been the focus of materials research.
In contrast to 3$d$ electron materials, in rare earth and actinide compounds, the spin-orbit interaction is much larger than the crystal electric field (CEF) interaction.
For Cerium-based rare-earth compounds, where only one electron is present in the 4$f$ shell,  the spin-orbit splitting between the two levels is of the order of 0.3 eV \cite{Patil.15,Cornut.72,Fujimori.03}. The degeneracy of the 4$f$ level is further reduced by crystal electric field effects according to the point-group symmetry of the lattice. For a tetragonal environment, {\itshape e.g.}, one expects that the $J=5/2$ and $J=7/2$ levels split into 3 and 4 doublets respectively \citep{Cornut.72}. The respective sizes of the CEF splittings typically lie between  several to few hundreds meV.  Various factors contribute to the CEF splittings in rare-earth and actinide-based heavy-fermion metals. It has long been recognized that
the size of the CEF splittings has a strong effect on the Kondo temperature ($T_K$), one of the important low-energy scales that characterize heavy fermions~\cite{Cornut.72,Kroha.03,Qiuyun.17,Zamani.17,Han.97}. Not only the effective hybridization between the local $f$ moments and the conduction bands, commonly only known qualitatively at best, is important but also the CEF splittings which together dynamically generate  $T_K$ as well as their mutual interdependence. The symmetry of the ground state CEF multiplet strongly affects  the $c$-$f$ electron hybridization \cite{Pagliuso.02}, and details of the CEF splittings also affect the competition between spin and orbital fluctuations \cite{Takimoto.02}. Nonetheless, the tools to reliably extract  information on the CEF configuration of rare earth and actinide intermetallics  is rather limited in number and scope \citep{Christianson.05,Christianson.04,Willers.10}.

The Ce$M$In$_5$ ($M$= Co, Rh, Ir) heavy-fermion compounds have attracted interest because they are stociometrically clean compounds possessing rich phase diagrams with competing ground states  and are good target materials to study {\itshape e.g.} how magnetism and superconductivity are related to each other \cite{Stewart.84, Si.14}. At ambient pressure, both CeCoIn$_5$ and CeIrIn$_5$ are superconductors with $T_c$ =2.3 K \cite{Petrovic.01} and 0.4 K \cite{Petrovic.01b}, respectively, while CeRhIn$_5$ is antiferromagnetic below $T_N$=3.8 K \cite{Mito.03}.

For CeIrIn$_5$, de Haas-van Alphen (dHvA) measurements found a Fermi surface volume expansion on going from LaIrIn$_5$ to CeIrIn$_5$ \cite{Shishido.02,Harrison.04}, which suggests that the $f$ electron delocalizes and participates at low temperature ($T$) in the Fermi surface of CeIrIn$_5$. A nuclear quadrupole resonance study of CeIrIn$_5$ lead to the conclusion that the 4$f$ electrons in CeIrIn$_5$ are much more itinerant than in the other known Ce-based heavy-fermion compounds \cite{Zheng.01}. This is in line with dynamical mean-field theory (DMFT) which predicts the formation of the heavy-fermion state in CeIrIn$_5$ at low temperature \cite{Shim.07, Choi.12, Haule.10}.
Angle-resolved photoemission spectroscopy (ARPES) measurements on CeIrIn$_5$, however, yield ambivalent results. Some claimed that the Ce 4$f$ electrons are nearly localized \cite{Fujimori.03}, while others directly observe a quasiparticle band in CeIrIn$_5$ \cite{Fujimori.06}. This ambiguity in understanding the low-$T$ properties of the $f$-electrons in CeIrIn$_5$ are at least in part rooted in the lack of a systematic electronic structure study of this important heavy-fermion compound.

In this paper, we provide this electronic structure study of CeIrIn$_5$ and trace the evolution of the $f$ spectral weight over an extended $T$ range. We find that the hybridization is stronger in CeIrIn$_5$ than CeCoIn$_5$ and the effects of the hybridization on the Fermi volume increase is much smaller than previous DMFT calculations \cite{Choi.12}. Importantly,  we are able to resolve the fine structure of both $4f_{5/2}^{1}$ and $4f_{7/2}^{1}$ and show how this fine structure contains information that relates to different contributions of the CEF splittings.

\begin{figure*}
\includegraphics[width=\textwidth]{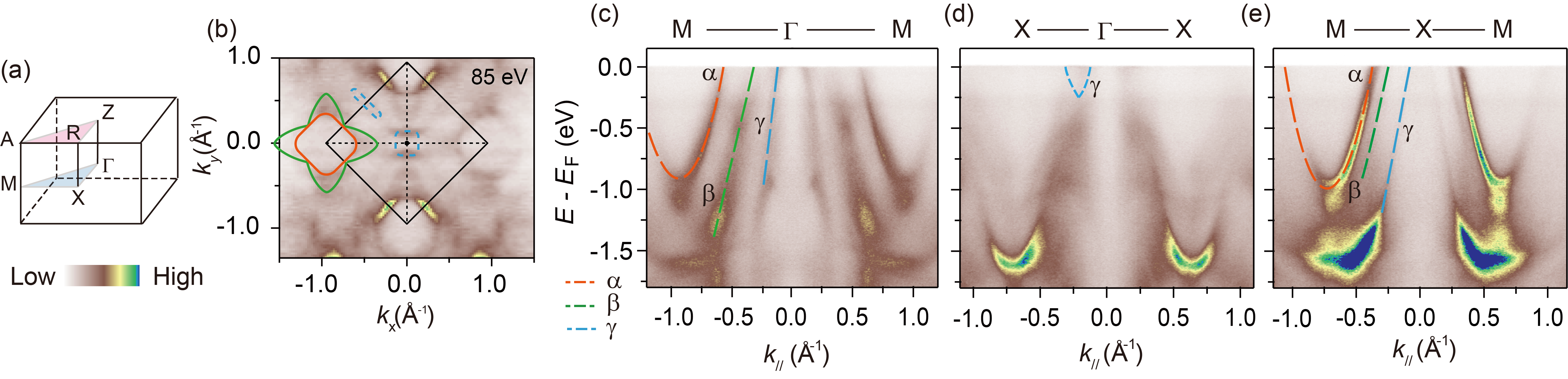}
\caption{Fermi surface and band structure of CeIrIn$_5$ taken with 85 eV photon at 12 K. (a) The Brillouin zone of CeIrIn$_5$, (b) Photoemission intensity map of CeIrIn$_5$ at $E_F$. The intensity is integrated over a window of ($E_F$-10 meV, $E_F$+10 meV). (c-e) Photoemission intensity distributions along (c) $\Gamma$-$M$, (d) $\Gamma$-$X$, and (e) $M$-$X$. }
\label{cut}
\end{figure*}
\begin{figure*}
\includegraphics[width=14cm]{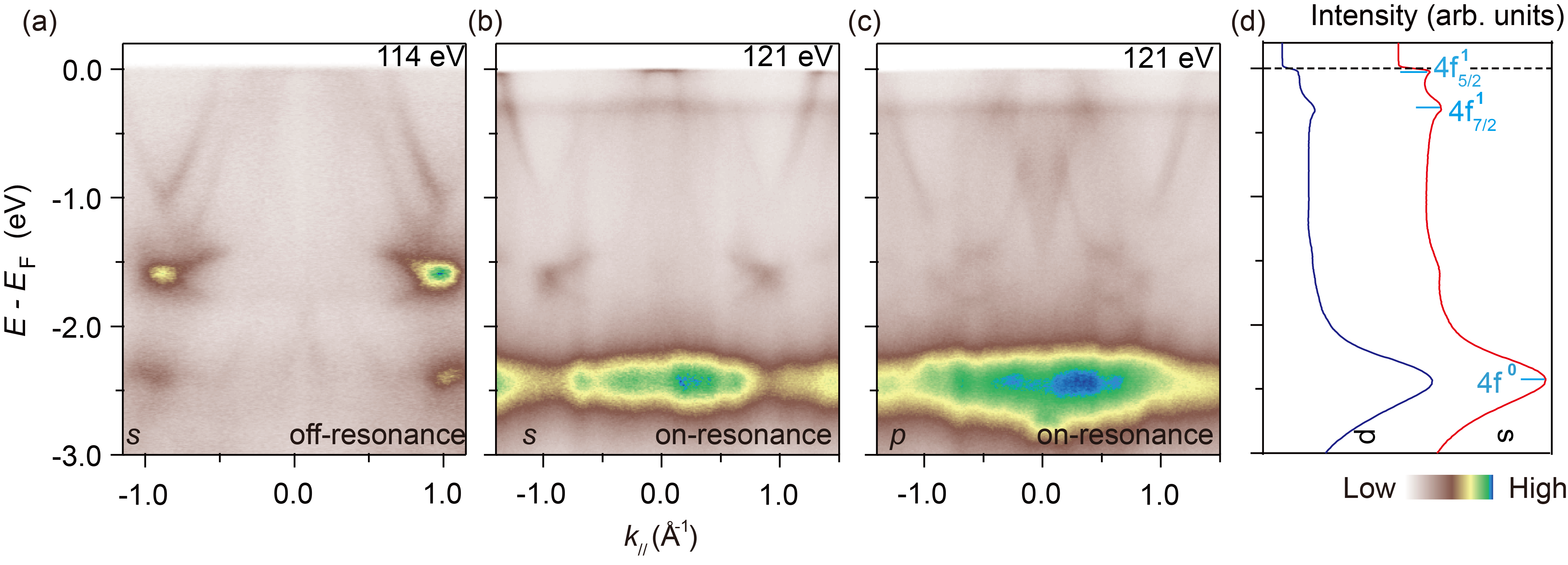}
\caption{Photoemission intensity distributions of CeIrIn$_5$  along $\Gamma$-M taken at 12 K with (a) off-resonance (114 eV)  $s$-polarized, (b) on-resonance (121~eV) $s$-polarized, and  (c)  on-resonance $p$-polarized photons.  (d) Angle-integrated EDCs of CeIrIn$_5$ taken with on-resonant energy; $f$-band positions are highlighted.
Momentum cuts taken with 121~eV photons cross (0,0,7.08$\frac{2\pi}{c}$), close to $\Gamma$, and are thus labeled $\Gamma$-M for simplicity.}
\label{band}
\end{figure*}

\begin{figure}
\includegraphics[width=\columnwidth]{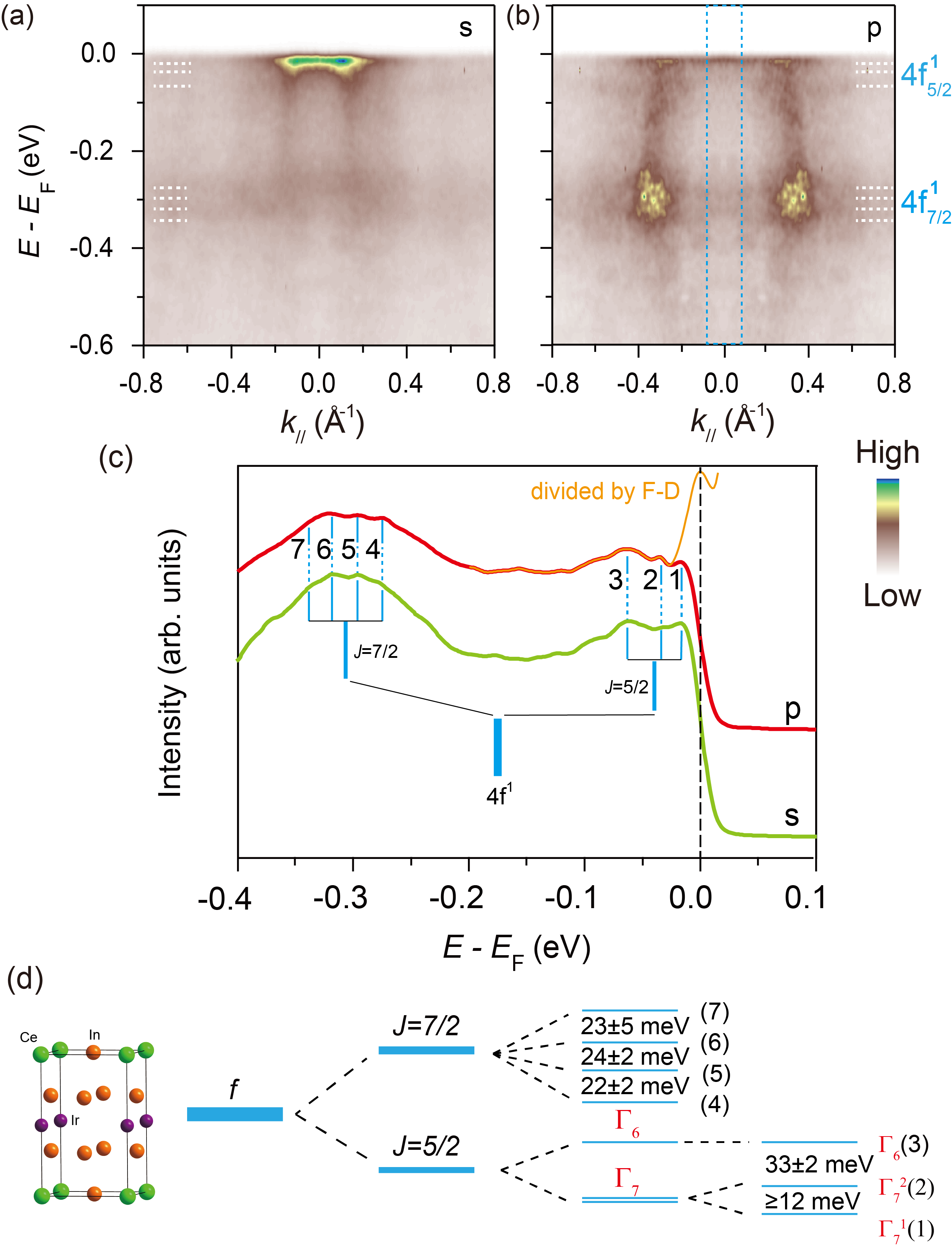}
\caption{Crystal electric field splittings of the 4$f$ states in CeIrIn$_5$. (a, b) Photoemission intensity plot along $\Gamma$-$M$ near $E_F$ taken with $s$-polarized photons (a) and $p$-polarized photons (b). (c) Integrated EDCs of CeIrIn$_5$ with $s$- and $p$- polarized photons. The integrated window has been marked with a blue block. The orange curve is the integrated EDCs of CeIrIn$_5$ with $p$- polarized photons after divided by the resolution-convoluted Fermi-Dirac distribution, from which the Kondo resonance peak above $E_F$ can be well identified. Since peak ``1" is influenced by the tail of the Kondo resonance, it is not clear whether it is a Kondo signature of the $\Gamma_7^{1}$ state. (d) Sketch of the CEF splittings of the multiplet $f$-states.}
\label{CEF}
\end{figure}
%
\section{II. Experimental details}
High quality single crystals of CeIrIn$_5$ were grown by an In self-flux method \cite{Petrovic.01b}.
All ARPES data presented here except those shown in Fig. \ref{band}  were performed at Beamline I05-ARPES of the Diamond Light Source equipped with a VG-Scienta R4000 electron analyzer. The typical angular resolution is $0.2^{\circ}$ and the overall energy resolution is better than 17 meV. Samples were cleaved $in$-$situ$ at 12~K and  below $9\times10^{-11}$~mbar.
The data in Fig. 2 were obtained at the ``Dreamline" beamline of the Shanghai Synchrotron Radiation Facility (SSRF) with a Scienta DA30 analyzer, and the vacuum was kept below $2\times10^{-10}$~mbar. The overall energy resolution was 20 meV. The samples were cleaved at 12 K.

\section{III. Results and discussions}

The Brillouin zone of CeIrIn$_5$ is sketched in Fig. 1(a) and the photoemission intensity map of CeIrIn$_5$ at 12 K with 85 eV photon energy is displayed in Fig. 1(b). The Fermi surface consists of one square-like Fermi pocket around the zone center which is part of the
$\gamma$ band, two Fermi pockets around the zone corner -- a flower-shaped $\beta$ and a squarelike $\alpha$ pocket -- and one narrow racetrack pocket extending to the middle of the zone boundary, which is also assigned to the $\gamma$ band. Detailed band dispersions derived from the photoemission intensity plots along several high-symmetry directions are shown in Figs. 1(c-e). Along $\Gamma$-$M$ in Fig. 1(c), three bands cross the Fermi level ($E_F$), which are assigned to $\alpha$, $\beta$, and $\gamma$, respectively. The hole-like $\gamma$ band encloses the $\Gamma$ point, forming the squarelike $\gamma$ Fermi pocket around the zone center; while the $\alpha$ band is parabolic-like with its bottom 0.95 eV below $E_F$. The Fermi energy crossings of the three bands can also be clearly observed along $M$-$X$, see Fig. 1(e). The bands crossing $E_F$ along $\Gamma$-$X$ in Fig. 1(d) are all from the $\gamma$ band, which shows strong $k_z$ dependence \cite{Qiuyun.17}.

\begin{figure*}
\includegraphics[width=\textwidth]{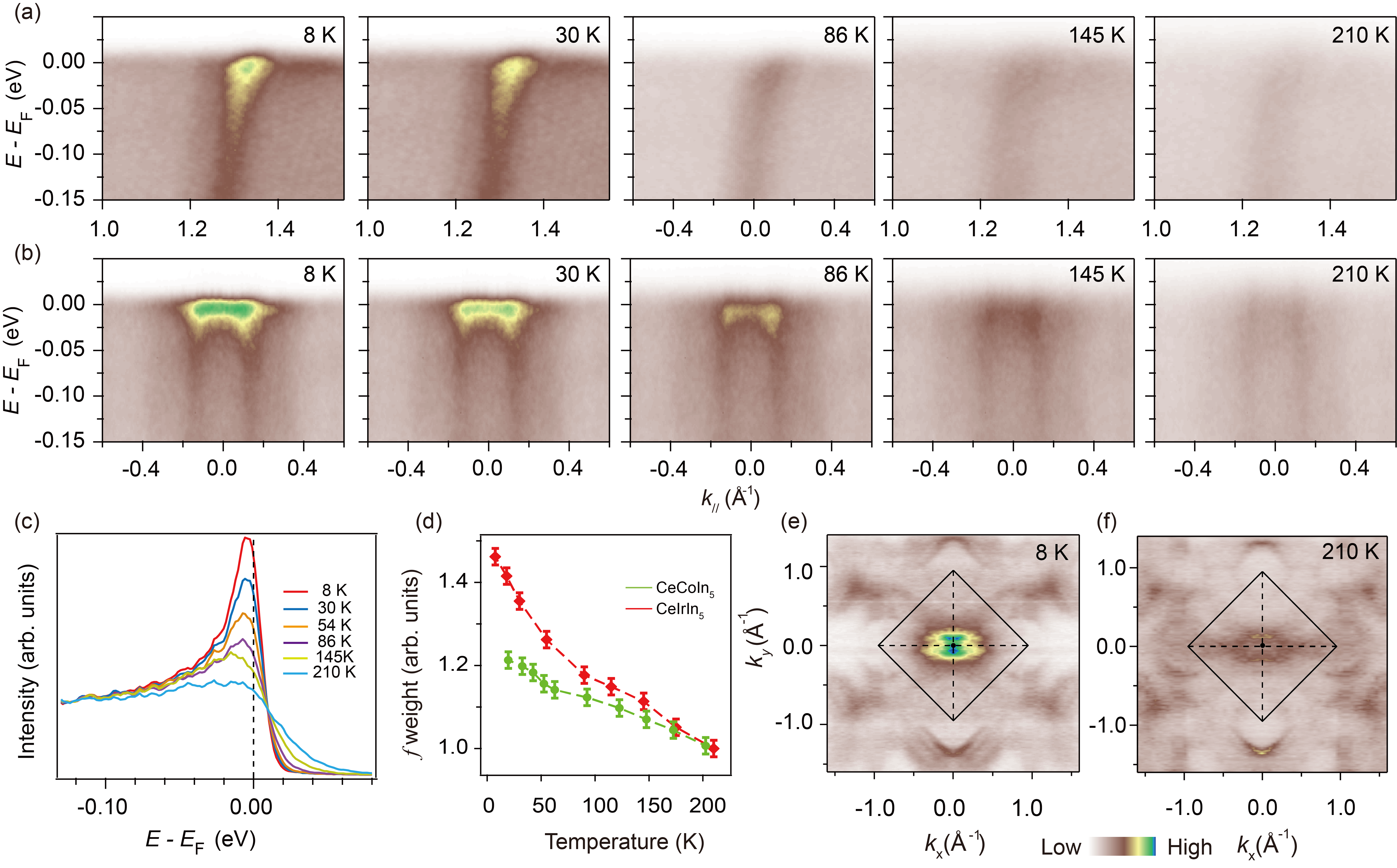}
\caption{$T$ evolution of the electronic structure of CeIrIn$_5$. (a) Zoomed-in ARPES data of the $\beta$ and $\gamma$ bands along $\Gamma$-$M$ at the $T$ indicated. (b) The same as (a), but for the $\alpha$ band.  (c) $T$ dependence of the EDCs at $\Gamma$. (d) $T$ dependence of the quasiparticle spectral weight in the vicinity of $\Gamma$ near $E_F$, integrated over [$E_F-$100~meV, $E_F+$10~meV] of CeIrIn$_5$ and CeCoIn$_5$ \cite{Qiuyun.17}. (e-f) Photoemission intensity map of CeIrIn$_5$ at $E_F$ at 12 K (e) and 210 K (f). The intensity is integrated over a window of ($E_F$-10 meV, $E_F$+10 meV).}
\label{tdep}
\end{figure*}

\begin{figure}
\includegraphics[width=\columnwidth]{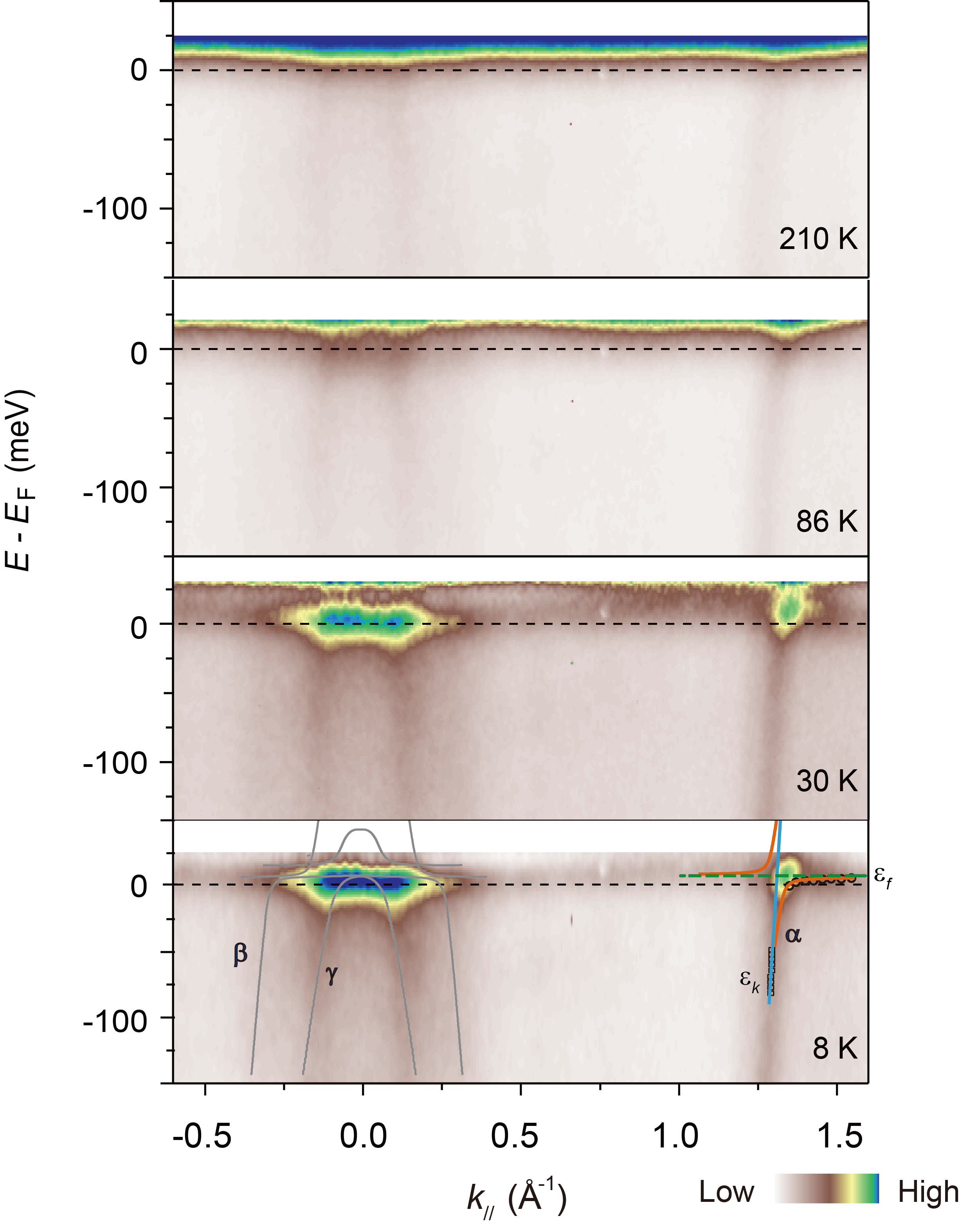}
\caption{Development of the heavy quasiparticle band in CeIrIn$_5$. Photoemission intensity plot along $\Gamma$-$M$ near $E_F$ taken at different temperatures, divided by the Resolution-convoluted Fermi-Dirac distribution. The spectra taken at 8 K has been fitted by the periodic Anderson Model. The gray solid line illustrates how the conduction $\beta$ and $\gamma$ bands hybridize with the $f$ bands. For the $\alpha$ band on the right side, we present a schematic diagram of the hybridization between $f$ electrons  ($\varepsilon_f$) and the $\alpha$ band ($\varepsilon_k$) close to $E_F$ under a periodic Anderson model. The orange curve is the hybridized band. Circles represent the position of the hybridized $f$ band obtained by tracking EDCs, squares represent the position of the conduction band from fitting MDCs. The blue and green line denotes the high-$T$ band dispersion and the position of the $f$ band, respectively. The left side of the parabolic $\alpha$ band is hardly observed due to matrix element effects.}
\label{CEFb}
\end{figure}

%
Resonant ARPES measurements were conducted at the Ce 4$d$--4$f$ transition to enhance the $f$-electron photoemission intensity.   A comparison of Figs.~2(a) and 2(b) illustrates the enhancement between the off-resonance (Fig. 2(a)) and on-resonance (Fig. 2(b)) photoemission intensities, respectively. Strongly dispersing bands dominate the off-resonance spectra, while Ce 4$f$ emission is enhanced in the on-resonance data, see also the integrated spectra in Fig. 2(d). Three nearly flat bands can be observed in the on-resonance data. The one at 2.3 eV binding energy with strong intensity is assigned to the initial $4f^0$ state, while those near $E_F$ and at 0.27 eV below $E_F$ are attributed to the $4f_{5/2}^{1}$  and its spin-orbit-split  $4f_{7/2}^{1}$ component, respectively. As it turns out, the $4f_{5/2}^{1}$ state is quite sensitive to the polarization of the light --- a significant enhancement is seen under $s$-polarized light compared to $p$ polarization (Fig.~2(c)). In contrast, the overall feature of the $4f_{7/2}^{1}$ and $4f^0$ states  appears insensitive to the polarization.


Figures 3(a) and 3(b) zoom into the vicinity of $E_F$, where a fine structure can be clearly identified. In particular, three peaks, located at -20, -32 and -65 meV, can be resolved which together form the $4f_{5/2}^{1}$ state, labeled 1, 2, 3 respectively in Fig.\ 3(c).
Surprisingly, we are also able to detect a fine structure of the $4f_{7/2}^{1}$ state which is comprised of four peaks. Three of them are located at -272, -294 and -318 meV (energies are measured with respect to $E_F$, {\itshape i.e.} $E_F=0$), while an additional shoulder peak at around -341 meV can also be identified, see Fig.\ 3(c). Since this shoulder peak is rather weak, exact energy position of this peak is yet to be determined. We also note that there is another shoulder peak about 90 meV away from feature 7. This may due to experimental statistical fluctuations, since the energy separation is too large for CEF splittings. To the best of our knowledge, the splitting of the $4f_{7/2}^{1}$  peak has not been detected before in any of the Ce-based heavy-electron materials (or analogously the $4f_{5/2}^{1}$ peak in Yb-based heavy-fermion compounds ).
The observation of the fine structure of the $4f_{7/2}^{1}$ peak at around -0.3 eV  establishes beyond doubt that it originates from Kondo scattering at the $J=7/2$ octet whose degeneracy is reduced to four doublets due to CEF effects. The absence of a clear $T$ dependence of this peak is related to a comparatively  high $T_K$ which results from  the relatively large number of contributing states, {\itshape i.e.}, the octet in absence of CEF effects.
CEF splittings of the $4f_{5/2}^{1}$ state have been directly observed in CeRh$_2$Si$_2$ by ARPES \cite{Patil.15}, and have been revealed in the Ce$M$In$_5$ compounds by inelastic neutron scattering measurements \cite{Christianson.04,Christianson.05,Willers.10}.
Interestingly, according to inelastic neutron scattering measurements on CeIrIn$_5$ \cite{Willers.10},  the energy splitting between the ground  ($\Gamma_7^{1}$)  and first excited state  ($\Gamma_7^{2}$) is 4 meV, while the ground state - second excited state ($\Gamma_6$) energy difference  is 28 meV.
In contrast, the energy separation revealed by our ARPES measurements is about 33 meV for the energy splitting between $\Gamma_7^{2}$ and $\Gamma_6$. It is less clear if the peak at position 1 in Fig.\ 3(c),  located at -20 meV, is likely due to the Fermi cutoff of the strong Kondo resonance above $E_F$, which overwhelms the $\Gamma_7^{1}$ state. As neither s- nor p- polarization indicate an additional peak at energies between 1 and 2, see Fig.\ 3(c), we are lead to conclude that the splitting between the $\Gamma_7^{1}$ and $\Gamma_7^{2}$ feature is at least 12 meV.

What could cause such a difference between the neutron scattering and the ARPES results?
One possibility is that the two techniques have different probing depth of the samples. In our previous soft x-ray ARPES studies on CeCoIn$_5$, we have demonstrated that ARPES data with 121 eV photons were still dominated by the bulk states \cite{Qiuyun.17}. Still, ARPES measurements mainly probes the surface, which might give different results from the bulk with neutron scattering.
Another possibility is that the CEF-derived Kondo satellites near $E_F$ develop dispersion due to a momentum-dependent hybridization, {\itshape i.e.}, the elements $V_{k,\delta}^m$ defined below.
This has been evident from previous ARPES measurements of the crystal-field splittings of the Kondo satellites in YbRh$_2$Si$_2$, in which energy dispersion in momentum space  has been detected \cite{Vyalikh.10}. Here, however, the aforementioned difference is observed at the $\Gamma$ point which rules out the momentum dispersion as its origin.
%
%

To address the possible origin of the observed difference,
we note that a minimal
model for CeIrIn$_5$ and related materials is
the multi-band multi-level periodic Anderson lattice model, given by
\begin{eqnarray}
\label{eq:Hamiltonian}
&&H=\sum_{i}H\big (\{f^\dagger\}_{i},\{f_{}^{}\}_i,\mathbf{R}_i\big ) \\
&&H\big (\{f^\dagger\},\{f_{}^{}\},\mathbf{R}_0\big )=\sum_{m,\sigma}\varepsilon^m_{} f^\dagger_{m\sigma}f^{}_{m\sigma}+
\sum_{\mathbf{k},\sigma,\delta}\varepsilon^\delta_{\mathbf{k}}c^\dagger_{\mathbf{k},\delta}c^{}_{\mathbf{k},\delta} \nonumber \\
&+&\frac{1}{2}\sum_{\stackrel{(m,\sigma)\neq}{ (m'\sigma')}}U_{mm'}^{\sigma \sigma'}f_{m\sigma}^{\dagger}f_{m'\sigma'}^{\dagger}f_{m'\sigma'}^{}f_{m\sigma}^{} \nonumber \\
&+&\sum_{\mathbf{k},\delta,m,\sigma}\Big(V_{\mathbf{k},\delta}^{m} f_{m\sigma}^{\dagger}c_{\mathbf{k}\delta,\sigma}^{}e^{i\mathbf{k}\cdot \mathbf{R}_0}+\mbox{h.c.}\Big),\nonumber
\end{eqnarray}
where $i$ labels the 4$f$  lattice sites located at positions $\mathbf{R}_i$ and $\{f^\dagger\}_{i}$ ($\{f_{}^{}\}_i,\mathbf{R}_i$) is the set of local $f$-electron creation (destruction) operators at site $i$. At each site $i$, the local operators carry each a set of indices $(m,\sigma)$ which distinguish the different CEF states: $m$ refers to the different doublets and $\sigma=\pm$  labels the states of each doublet.  The index $\delta$ distinguishes the different conduction bands.
$c^\dagger_{\mathbf{k},\delta,\sigma}$ creates a conduction electron of lattice momentum $\mathbf{k}$ and spin projection $\sigma$ in the band $\delta$. $U_{mm'}^{\sigma \sigma'}$ is the local Coulomb matrix element and $V_{\mathbf{k},\delta}^{m}$ denotes the hybridization strength between corresponding states.
In order to discuss the CEF splittings, it is sufficient to consider  a single site $i=i_0$ ($R_0=0$) for simplicity. We will refer to the set $\{\varepsilon^m\}$ as {\itshape bare} CEF levels. The origin of this contribution to different CEF levels is  the  charge distribution in the vicinity of site $i_0$. From Eq.~(\ref{eq:Hamiltonian}) it follows that the $\varepsilon^m$ and  the resulting CEF splittings $\Delta_{mn}=\varepsilon^m-\varepsilon^n$  are independent of  $\{V_{\mathbf{k},\delta}^{m}\}$ and $\{U_{mm'}^{\sigma \sigma'}\}$.

Analyzing the model system Eq.\ (\ref{eq:Hamiltonian}) for  a single site $i=i_0$ ($R_0=0$) shows that the set $\{\varepsilon^m\}$ of bare CEF levels
is strongly renormalized in  heavy-fermion metals and the amount of renormalization will depend both on $\{V_{\mathbf{k},\delta}^{m}\}$ and $\{U_{mm'}^{\sigma \sigma'}\}$ as well as details of the conduction electron bands~\cite{Zamani.17}.
One way of approximately calculating the {\itshape fully renormalized} set of energies $\{\tilde{\varepsilon}^m_{}\}$ is based on the perturbative RG method of \cite{Jefferson.77,Zamani.17}.
For $M=1$, $U_{mm'}^{}\rightarrow \infty$ and a single conduction band  of width $2D$ {\itshape e.g.} one finds $\tilde{\varepsilon}_{1}=\varepsilon_{1}+ |V_1|^2/2D\ln\big(\tilde{\varepsilon}_1/D\big)$ in the so-called wideband limit, where $\varepsilon_1\ll D$.
What is measured by XPS or neutron spectroscopy is the  renormalized set of energies $\{\tilde{\varepsilon}^m_{}\}$ of the $4f^{0}$ multiplet in the fully interacting system of Eq.\ (\ref{eq:Hamiltonian}).
In contrast, ARPES allows one to determine the position of CEF-derived Kondo satellite peaks near $E_F$. The energy spacings between the Kondo satellites are again the result of the full many-body problem in the fully interacting system. Yet, they are in principle different from those obtained by XPS and neutron spectroscopy. An approximate method to obtain the spacings between the Kondo satellites is the slave boson mean field theory which for the present model predicts that a set of Kondo resonances appear both above and below $E_F$. Applying the slave boson mean field theory to Eq.\ (\ref{eq:Hamiltonian}) with  $U_{mm'}^{\sigma \sigma'}\rightarrow \infty$ one finds, that the spacings between the Kondo satellites are essentially those between the unrenormalized set of CEF levels $\{\varepsilon^m\}$~\cite{Kroha.03,Zamani.17}.
Thus, we can conclude that the observed differences between the CEF splittings from ARPES, XPS and neutron scattering  are due to many-body effects. At the level of rigor discussed above, the splittings of the Kondo satellite peaks reflect  the {\itshape bare} CEF splittings which enables us to disentangle the two factors contributing  to CEF splittings occurring in heavy-electron metals.
From our analysis and the observation that among the Ce$M$In$_5$ family, only CeIrIn$_5$ shows a detectable difference.  We can thus conclude that one of the $V_{m}$ for the $4f_{5/2}$ state dominates over the other two. This then would place CeIrIn$_5$ much closer into the mixed valence regime, as compared to CeCoIn$_5$ or CeRhIn$_5$.  This naturally explains why CeIrIn$_5$ has the smallest mass enhancement of the Ce$M$In$_5$ family.

To investigate the evolution of the $f$ electrons as a function of $T$, we performed $T$-dependent resonant ARPES with 121 eV photons along $\Gamma M$. Figs. 4(a) and 4(b) display the evolution of the $\alpha$ band (a), and $\beta$ and $\gamma$ bands (b). At high $T$, all three conduction bands show fast dispersive features. Upon decreasing $T$, the bands start bending and a weakly dispersive $f$-electron feature gradually emerges with increasing weight near $E_F$, which is particularly pronounced near $\Gamma$. The $T$ evolution is further illustrated by the EDCs around $\Gamma$ in Fig. 4(c).  Our results demonstrate that the heavy band formation begins at $T$ much higher than the coherence temperature $T_{coh}$, which for CeIrIn$_5$   $\thicksim$ 50 K \cite{Choi.12}, as can be inferred from Figs. 4(a) and 4(b). The $f$ spectral weight near $E_F$ is already discernible at around 145 K, and weakly dispersive hybridized band become discernible at around 90 K.

In Fig. 4(d) we present a comparison of the $T$-dependence of  $f$ spectral weight between CeIrIn$_5$ and its sister compound CeCoIn$_5$. For comparison, the spectral weight has been normalized such that it coincides for both  compounds at 200 K, where the effect of hybridization is relatively small.  From Fig. 4(d), it is clear that the $f$ spectral weight in CeIrIn$_5$ increases much faster upon lowering $T$ than  in CeCoIn$_5$, and the larger $f$ spectral weight at low $T$ as compared to that  found in CeIrIn$_5$ at the same $T$, is indicative of a stronger hybridization in CeIrIn$_5$.  This is consistent with DMFT calculations, which find that CeIrIn$_5$ has the largest quasiparticle peak in the Ce$M$In$_5$ family \cite{Haule.10}. However, the observed stronger $c$-$f$ electron hybridization in CeIrIn$_5$ from our results is different from previous ARPES data, which suggests that the 4$f$ electrons are dominated by the localized character with small itinerant component \cite{Fujimori.06}. This difference may be due to the fact that the $4f_{5/2}^{1}$ state is quite sensitive to the polarization of the light---a significant enhancement is seen under $s$-polarization light compared with $p$ polarization.

Figures. 4(e) and 4(f) display the photoemission intensity maps for CeIrIn$_5$ taken at 8 K and 210 K. At low $T$ the $f$-electron spectral weight is considerably enhanced near $\Gamma$, indicating that the $f$-electrons participate in the electronic properties and the system forms a {\itshape large} Fermi surface, while at high $T$ the Fermi surface is {\itshape small}. Interestingly, we find that the effect of the hybridization on the Fermi volume increase is much smaller than predicted by DMFT calculations, which suggest a dramatic change of the Fermi surface topology at low $T$ \cite{Choi.12}.

Fig. 5 displays the spectra along $\Gamma$-$M$ taken at different temperatures after dividing by the resolution-convoluted Fermi-Dirac distribution, from which the formation of the heavy hybridized bands at low temperature can be more clearly read off. At high temperature, three conduction bands all show linear dispersion and no obvious redistribution of the $f$ spectral weight can be observed.
At low temperature, the hybridization of the two bands with the $f$ bands causes the redistribution of the $f$ spectral weight and forms a weakly dispersive band near $\Gamma$. The $f$ spectral weight is significantly enhanced to the `inside' of the two bands. While for the $\alpha$ band, one can clearly observe the dispersion of the Kondo resonance peak, and it can be well explained by the hybridized band picture based on the mean field theory for the periodic Anderson model (with $M=1$) \cite{Hewson.93}, as presented on the right side of the spectrum taken at 8 K of Fig.\ 5, in which the energy dispersion is given by
$E^{\pm}=\frac{\varepsilon_f+\varepsilon(k)\pm\sqrt{(\varepsilon_f-\varepsilon(k))^2+4|V_k|^2}}{2}$,
where $\varepsilon_f$ is the single ($M=1$) renormalized $f$-level energy, $\varepsilon_k$ is the conduction-band dispersion at high temperatures, and  $V_k$ is the renormalized hybridization \cite{Hewson.93,Im.08}. A fit to this model gives $\varepsilon_f$=1 meV, and $V_k$=$18\pm5$~meV for the $\alpha$ band, corresponding to a direct gap of 36 meV for the $\alpha$ band, a little larger than that of 30 meV found in CeCoIn$_5$ \cite{Qiuyun.17}. The larger hybridization gap is consistent with the stronger hybridization found in CeIrIn$_5$ as compared to CeCoIn$_5$.


\section{V. Conclusions}
To summarize, we provide the electronic structure study of the heavy-fermion superconductor CeIrIn$_5$ in a wide temperature range. We show how the localized $f$ moments evolute into the heavy-fermion state starting from a much higher temperatures than $T_{coh}$. The crystal electric field splittings of both $4f_{5/2}^{1}$ and $4f_{7/2}^{1}$ states have been directly observed for the first time in heavy-fermion compounds. We also addressed the difference of the CEF splittings inferred from ARPES vs that from other spectroscopy methods and have shown that this allows us to disentangle different contributions to the CEF splittings. Moreover, we find that the hybridization between the $f$  and conduction electrons is stronger in CeIrIn$_5$ than in CeCoIn$_5$. Our findings should prove  essential for a complete microscopic understanding of the intricate phase diagrams of
the Ce 115 compounds and related heavy-fermion systems.


\section{Acknowledgments}

S. K. acknowledges helpful discussions with F.\ Zamani, G.\ Zwicknagl, and J.\ Kroha.
This work is supported in part by the National Key Research and Development Program of China (Grant No.\ 2016YFA0300200 and No. 2017YFA0303104), the National Science Foundation of China ((Grants No. 11504342, No. 11474060, No. 11774320, No. 11474250, and No. U1630248),
Science Challenge Project (No.\ TZ2016004), Diamond Light Source for time on beamline I05 under Proposal No.\ SI11914. Some preliminary data were taken at the National Synchrotron Radiation Laboratory (NSRL).


\end{document}